\documentclass[12pt]{iopart}

\usepackage{graphicx}
\usepackage{amssymb}
\begin{document}

\title{Correlation between concurrence and mutual information}

\author{Yong Kwon$^1$, Seung Ki Baek$^2$, Jaegon Um$^{3\ast}$}

\address{$^1$ Department of Physics, Pukyong National University, Busan 48513, Korea}
\address{$^2$ Department of Scientific Computing, Pukyong National University, Busan 48513, Korea}
\address{$^3$ Department of Physics, Pohang University of Science and
Technology, Pohang 37673, Korea}
\ead{$^\ast$slung@postech.ac.kr}
\vspace{10pt}
\begin{indented}
\item[]March 2022
\end{indented}

\begin{abstract}
We investigate a two-qubit system to understand the relationship between concurrence and mutual information, where the former determines the amount of quantum entanglement, whereas the latter is
its classical residue after performing local projective measurement.
For a given ensemble of random pure states,
in which the values of concurrence are uniformly distributed, 
we calculate the joint probability of concurrence and mutual information. 
Although zero mutual information is the most probable in the uniform ensemble, 
we find positive correlation between the classical information and 
concurrence.
This result suggests that destructive measurement
of classical information can be used to assess the amount of quantum information.
\end{abstract}
%
%
%
\maketitle
%
%

\section{Introduction}
\label{sec:intro}

Quantum entanglement (QE) is a distinct feature of quantum systems~\cite{nielsen2000quantum}. Its quantitative measurement
requires full information of a wavefunction, and
several methods to measure wavefunctions have been
proposed~\cite{james2001measurement,lundeen2011direct,thekkadath2016direct,lundeen2005practical,brodutch2016nonlocal,pan2019direct}.
Mutual information (MI) can be regarded as a classical counterpart of quantum
entanglement. This quantity has widely been
used to determine correlation between subsystems~\cite{sagawa2010generalized,lestas2010fundamental,lau2013information,muller2013mutual}. It is also obtainable in a pure quantum state 
after local projective measurement which yields
a classical probability distribution with respect to the measurement basis.
Post-measurement
MI refers only to the diagonal part of a density operator;
this restriction implies that in general a part of the information content in QE will be lost
when it is converted to MI by
local projective measurement. Indeed, the following inequality has been proven for a bipartite system in a pure state:
\begin{equation}
    I \leq E,
    \label{eq:inequality}
\end{equation}
where $I$ is
post-measurement MI, and $E$ is von~Neumann entanglement entropy~\cite{um2012entanglement}.
The common wisdom is that MI is not a reliable measure of QE. For example, let us consider a two-qubit system given 
by $\vert \psi \rangle = \left( \vert 00\rangle + \vert 01\rangle 
- \vert 10\rangle + \vert 11 \rangle \right )/ 2$. Although the qubits are maximally entangled, projective measurement in the basis of $ \{\vert 00 \rangle, 
\vert 01 \rangle, \vert 10\rangle, \vert 11 \rangle \}$ fails to detect the entanglement, because one obtains uniform probability distribution with no classical correlation between the qubits, i.e., $I=0$.
Still, MI may sometimes serve as an indicator of QE~\cite{um2012entanglement}, as demonstrated by the scaling behavior in the quantum Ising chain at the critical point~\cite{calabrese2004entanglement}.

\begin{figure}
\centering
\includegraphics[width=0.80\textwidth]{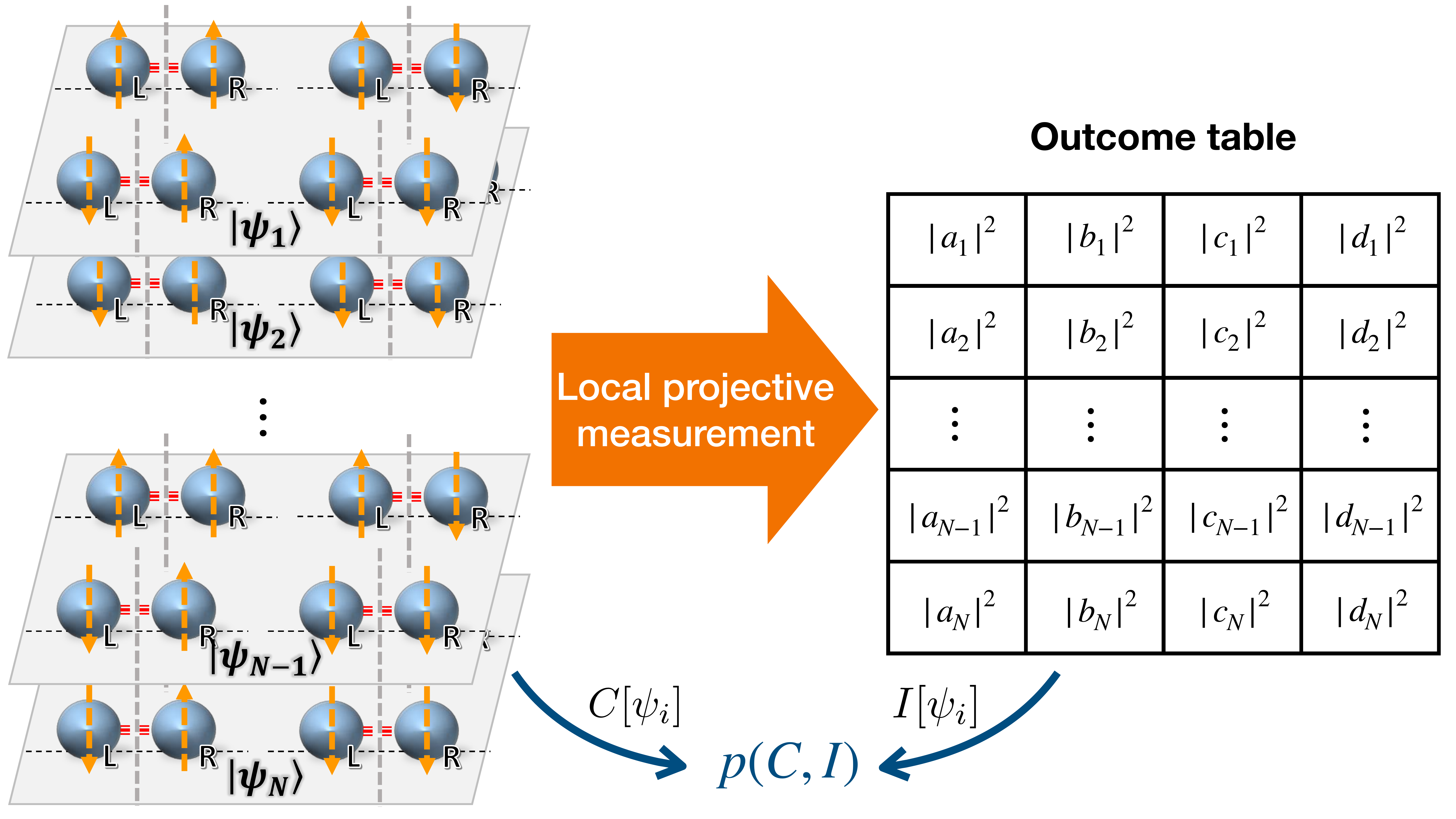} \\
\caption{Schematic diagram of this work.
We prepare an ensemble of $N$ pure states, 
each of which is written as
$\vert \psi_i \rangle = a_i \vert 00 \rangle + b_i \vert 01 \rangle 
+ c_i \vert 10 \rangle + d_i \vert 11 \rangle$.
Concurrence $C[\psi_i]$ is directly calculated from the wave function.
MI $I[\psi_i]$ is calculated from the
probability distribution
among the four configurations in $\vert \psi_i \rangle$, i.e., 
$\left( |a_i|^2, |b_i|^2, |c_i|^2, |d_i|^2 \right)$,
as obtained by using local projective measurement.
By combining these two quantities, we construct their joint PDF $p(C,I)$.
}
    \label{fig:scheme}
\end{figure}

In this work, we address the correlation 
between QE and MI by considering 
uniformly random two-qubit pure states (Fig.~\ref{fig:scheme}).
For the sake of analytic convenience, we will work with concurrence $C$ (defined below) as a measure of entanglement~\cite{hill1997entanglement} because
$E$ is explicitly written as a monotonically increasing function of $C$ in a pure state of two qubits~\cite{wootters1998entanglement}.
From the joint probability density function (PDF) of $I$ and $C$, we observe
positive correlation between them. We also provide an analytic expression for the most probable value of $C$ when $I$ is given.
For a given ensemble of random pure states,
one can thus infer the amount of QE
by applying local projective measurement to a randomly selected state.
It suggests how one can estimate the amount of quantum information through
a statistical method, which is destructive but relatively simple to implement.

This work is organized as follows:
In Sec.~\ref{sec:local} we present our observables and the measurement scheme to obtain MI. We also define `concurrence'. In Sec.~\ref{sec:joint}, we observe positive correlation between
concurrence and MI in random pure states by 
computing the joint PDF numerically. In Sec.~\ref{sec:partial}, we analytically derive the most probable value of $C$ for given $I$. In Sec.~\ref{sec:summary}, we summarize this work.

\section{Observables}
\label{sec:local}

\subsection{Mutual information}
Let us consider a pure state composed of two qubits,
`L' and `R'. The wave function is written as
\begin{equation}
\vert \psi \rangle = a \vert 00 \rangle + b \vert 01 \rangle 
+ c \vert 10 \rangle + d \vert 11 \rangle \,,
\label{eq:psi}
\end{equation}
where each base ket indicates the states of L and R. The coefficients are complex numbers, and
satisfy $|a|^2 + |b|^2 + |c|^2 + |d|^2=1$ because of the normalization condition.

Local projective measurement is implemented as follows:
We define the following projection operators:
\begin{eqnarray}
\Pi_1 &\equiv& \vert 00 \rangle \langle 00\vert \,, \,\,\,\,\,\,\, 
\Pi_2 \equiv \vert 01 \rangle \langle 01\vert \,, \nonumber\\ 
\Pi_3 &\equiv& \vert 10 \rangle \langle 10\vert \,, \,\,\,\,\,\,\,
\Pi_4 \equiv \vert 11 \rangle \langle 11\vert \,, 
\end{eqnarray}
to measure classical configurations of the system.
The corresponding outcome for configuration $m$ is obtained 
as $P_m = {\rm tr}\, \Pi_m \rho$,
where $\rho \equiv \vert \psi \rangle \langle \psi \vert$ is the density operator.
Therefore, when applied to Eq.~\eref{eq:psi}, the measurement outcomes are
$P_1[\psi]= |a|^2$, $P_2[\psi]= |b|^2$, $P_3[\psi]= |c|^2$, 
and $P_4[\psi]= |d|^2$.
The Shannon entropy of the total system is
\begin{equation}
H_{\rm tot} = -\sum_{m=1}^4 P_m\, \log_2 P_m \,, \\
\end{equation}
and those of the subsystems are
\begin{eqnarray}
H_{\rm L} &=& - \left(P_1 + P_2 \right) \log_2 \left(P_1 + P_2 \right)
- \left(P_3 + P_4 \right) \log_2 \left(P_3 + P_4 \right)  , \label{eq:hl}\\
H_{\rm R} &=& - \left(P_1 + P_3 \right) \log_2 \left(P_1 + P_3 \right)
- \left(P_2 + P_4 \right) \log_2 \left(P_2 + P_4 \right)  \,.
\end{eqnarray}
We then obtain post-measurement MI
$I[\psi]$ defined as~\cite{cover2006elements}
\begin{equation}
I[\psi] = H_{\rm L}[\psi] + H_{\rm R}[\psi] - H_{\rm tot}[\psi]\,.
\label{eq:I}
\end{equation}
For example, the Bell state
\begin{equation}
    \vert \phi_1 \rangle \equiv \left( \vert 00 \rangle + \vert 11\rangle
\right) /\sqrt{2}
\label{eq:bell}
\end{equation}
yields $I[\phi_1]=1$ because
$H_{\rm tot}[\phi_1]=H_{\rm L}[\phi_1] = H_{\rm R}[\phi_1] = 1$. Similarly, we find the same result for
\begin{equation}
    \vert \phi_2 \rangle \equiv \left( \vert 01 \rangle - \vert 10\rangle \right) /\sqrt{2}.
    \label{eq:phi2}
\end{equation}
However, their superposition
\begin{equation}
\vert \phi_3 \rangle  \equiv \left( \vert \phi_1 \rangle + \vert \phi_2 \rangle \right)/\sqrt{2}
\label{eq:phi3}
\end{equation}
has $I[\phi_3]=0$ because
$H_{\rm tot}[\phi_3]=2$ and $H_{\rm L}[\phi_3] = H_{\rm R}[\phi_3] = 1$.

\subsection{Concurrence}
\label{sec:concurrence}
For a pure state, QE between two sectors can be quantified
by the von~Neumann entropy of a subsystem~\cite{bennett1996concentrating}.
If we consider $\vert \psi \rangle$ in Eq.~\eref{eq:psi}, QE between 
L and R is given by
\begin{equation}
E[\psi] = -{\rm tr}\, \rho_{\rm L} \log_2 \rho_{\rm L} =  
-{\rm tr}\, \rho_{\rm R} \log_2 \rho_{\rm R} \,,
\label{eq:E}
\end{equation}
where $\rho_{\rm L} \equiv {\rm tr}_{\rm R} \vert \psi \rangle \langle \psi \vert$
and $\rho_{\rm R} \equiv {\rm tr}_{\rm L} \vert \psi \rangle \langle \psi \vert$
are the density operators of L and R, respectively.
In the above example [Eqs.~\eref{eq:bell} to \eref{eq:phi3}], obviously
$E[\phi_1]=E[\phi_2]=E[\phi_3]=1$. Equation~\eref{eq:E} correctly detects quantum correlation between the subsystems in $\vert \phi_3 \rangle$ [Eq.~\eref{eq:phi3}], whereas MI does not. Of course,
this result satisfies the more general inequality in Eq.~\eref{eq:inequality}.

For $\vert \psi \rangle$ in Eq.~\eref{eq:psi}, the von~Neumann entropy is written as
\begin{equation}
E[\psi]=E\left( C[\psi] \right)=-x \log_2 x - (1-x) \log_2 (1-x) \,,
\label{eq:E_C}
\end{equation}
where
\begin{eqnarray}
x(C) &\equiv& \frac{1 + \sqrt{1-C^2} }{2} \,,\\
C[\psi] &\equiv& 2 \left| ad - bc \right| \,.
\label{eq:C}
\end{eqnarray}
Equation~\eref{eq:C} defines concurrence throughout this work.
Both $E$ and $C$ take values within the unit interval $[0,1]$, and $E$ is a monotonically increasing function of $C$. The end points are $E(C=0)=0$ and $E(C=1)=1$, and zero entanglement in the product space thus corresponds to $ad=bc$.

\section{Result}
\label{sec:joint}

\begin{figure}
\centering
\includegraphics[width=0.9\textwidth]{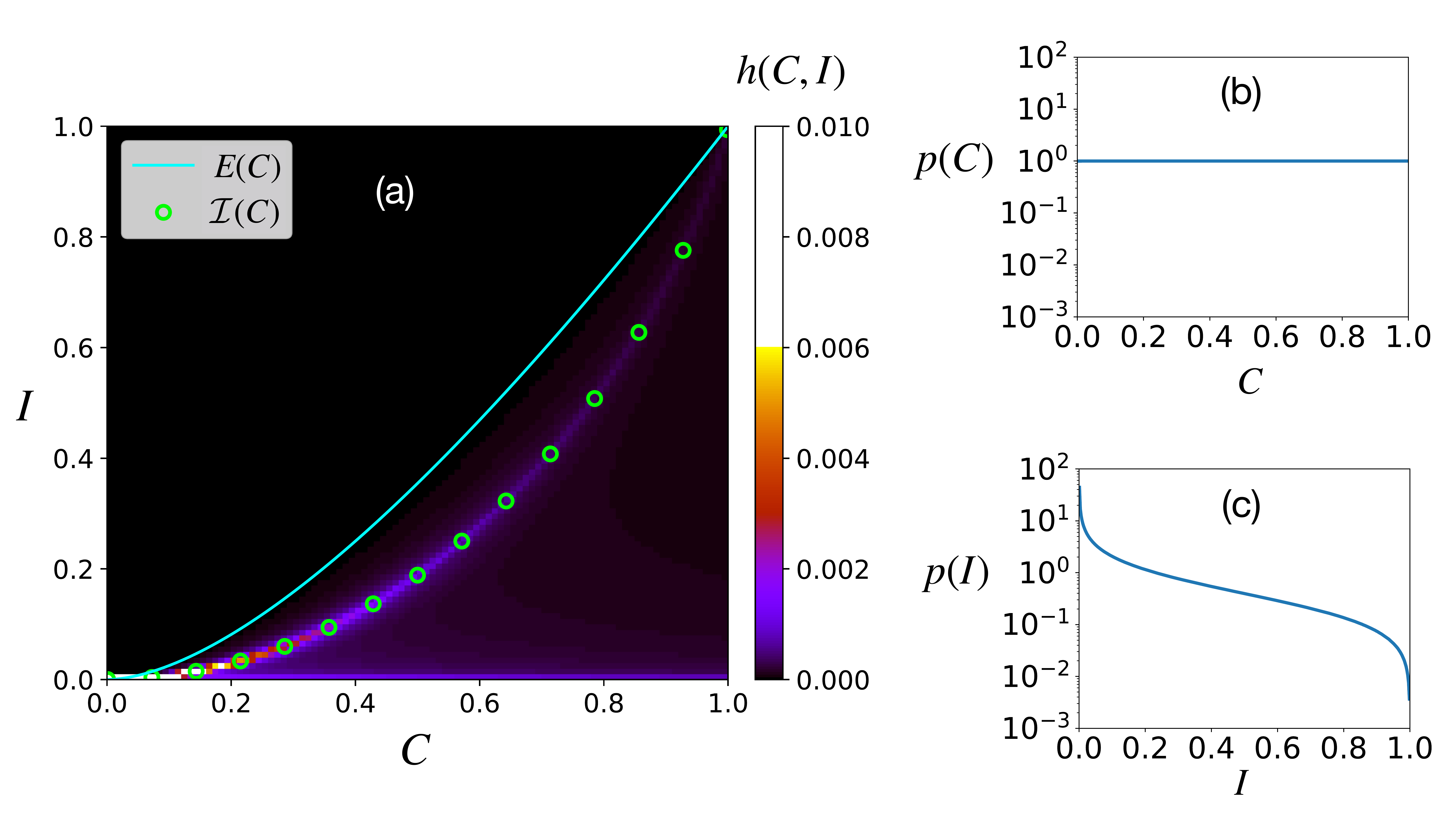}
\caption{(a) Histogram $h(C,I) \equiv p(C,I) \Delta C \Delta I$
with $\Delta C = \Delta I = 10^{-2}$, obtained from an ensemble of 
$N=2\times 10^{9}$ pure states,
where the coefficients are real numbers uniformly chosen at random under the normalization condition, i.e., $a_i^2 +b_i^2 +c_i^2 +d_i^2 =1$.
Solid line: General upper bound in Eq.~\eref{eq:inequality}. The joint probability is concentrated on a narrow band below the solid line, and its functional form is described by Eq.~\eref{eq:bound} in Sec.~\ref{sec:partial} (open circles).
(b) and (c): Marginal PDFs for concurrence ($p(C)$) and MI ($p(I)$), respectively. The simulation parameters are the same as
in (a), except that $\Delta C = \Delta I = 2.5\times 10^{-3}$.
}
\label{fig:PDF}
\end{figure}

We express the complex coefficients in Eq.~\eref{eq:psi} as
\begin{equation}
a=|a|e^{{\rm i} \theta_a} \,,
b=|b|e^{{\rm i} \theta_b} \,,
c=|c|e^{{\rm i} \theta_c} \,,
d=|d|e^{{\rm i} \theta_d} \,
\label{eq:phaseangles}
\end{equation}
where each phase takes a uniform random value within $[0,2\pi)$.
Concurrence is then rewritten as
\begin{equation}
C[\psi] = 2 \sqrt{ |ad|^2 + |bc|^2 - 2|abcd| \cos \theta }\,,
\label{eq:C_theta}
\end{equation}
where $\theta \equiv \theta_a + \theta_d - \theta_b -\theta_c$.
Assume that the coefficients are randomly chosen under the condition that $|a|^2 + |b|^2 + |c|^2 + |d|^2 = 1$. The resulting angle $\theta$ will again be random, drawn from a uniform PDF denoted as $u(\theta)$.
The most probable value of $v \equiv \cos\theta$ is either $+1$ or $-1$ because its distribution $f(v) = u(\theta) \left| d\theta / dv \right|$ has a peak when $\theta$ equals an integer multiple of $\pi$, at which $dv / d\theta = -\sin \theta = 0$. Around the peak positions, concurrence can be approximated as
\begin{equation}
C[\psi] \approx 2 \sqrt{ |ad|^2 + |bc|^2 \pm 2|abcd| } = 2 \left| |ad| \pm |bc| \right|\,,
\label{eq:C_theta2}
\end{equation}
which implies that the phases are mostly irrelevant.
For this reason, we henceforth focus on real coefficients. In other words, among an ensemble of $N$ pure quantum states, the $i$th wave function is now described by $\vert \psi_i \rangle = a_i \vert 00 \rangle + b_i \vert 01 \rangle 
+ c_i \vert 10\rangle + d_i \vert 11 \rangle$, for which all the coefficients are real.
The effective Hilbert space is thus reduced to the unit 3-sphere $S^3 \equiv \left\{ \vec{r} \in \mathbb{R}^4 : || \vec{r} || = 1 \right\}$.

For every $\vert \psi_i \rangle$, we calculate $C[\psi_i]$ and $I[\psi_i]$ by using Eqs.~\eref{eq:I} and \eref{eq:C}. The joint PDF [Fig.~\ref{fig:PDF}(a)] for $C$ and $I$ is then obtained as
\begin{equation}
p(C,I) \equiv N^{-1} \sum^N_{i=1}
\delta\left( C-C[\psi_i] \right)
\,\delta \left(I - I[\psi_i] \right)\,,
\label{eq:joint}
\end{equation}
where $\delta$ means the Dirac delta function.
We also obtain two marginal PDFs that are defined as
$p(C) \equiv \int dI~ p(C,I)$ [Fig.~\ref{fig:PDF}(b)] and $p(I) \equiv \int dC~ p(C,I)$ [Fig.~\ref{fig:PDF}(c)].
Every sample in our ensemble confirms the general upper bound in Eq.~\eref{eq:inequality}, i.e., $I \le E(C)$.
The important point is that the joint PDF is not uniform but has concentrated regions. The narrow band below the solid line in Fig.~\ref{fig:PDF}(a) clearly demonstrates nontrivial correlation between $C$ and $I$. We denote functional form of this correlation as $I=\mathcal{I}(C)$, and
will derive it analytically in Sec.~\ref{sec:partial}.

\begin{figure}[h]
\centering
\includegraphics[width=\textwidth]{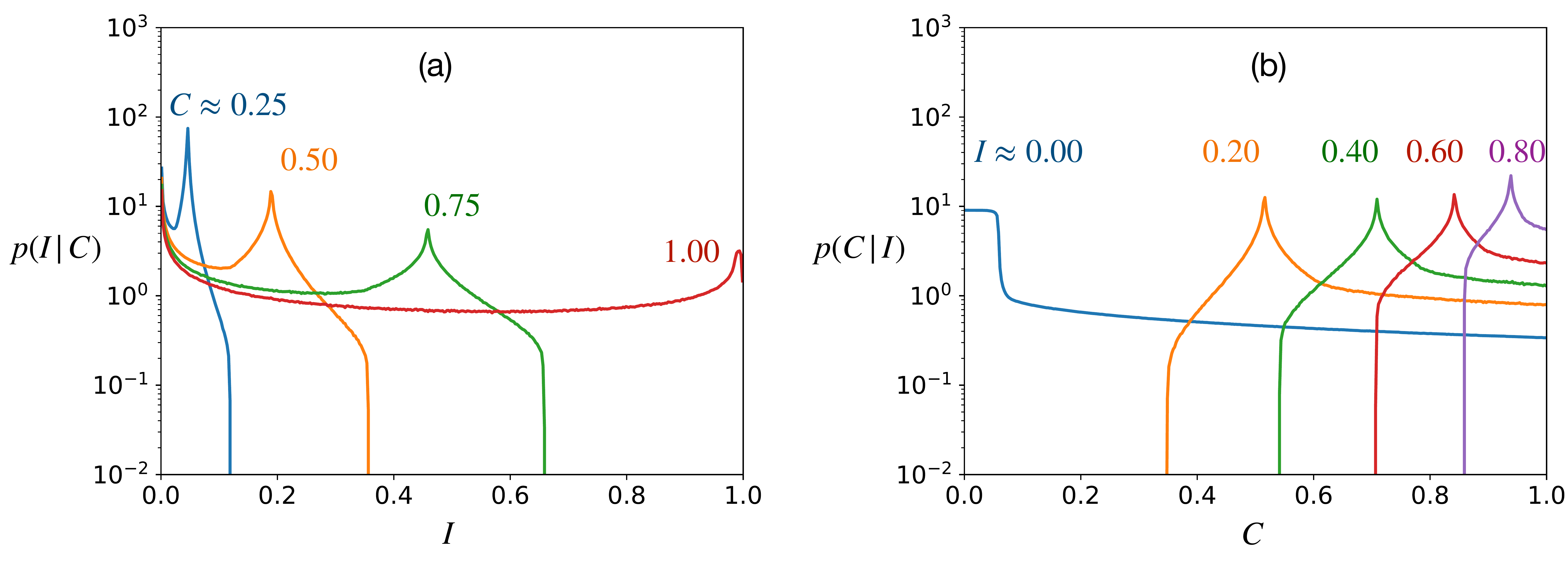} 
\caption{Conditional PDF's obtained from Fig.~\ref{fig:PDF}.
(a) $p(I|C)$ as $C$ varies. Each curve has two peaks, one at $I=0$ and the other at $I>0$.
(b) $p(C|I)$ for various values of $I$. Every curve has a single peak, from which the most probable value of $C$ can be predicted when $I$ is given by experiments. For the leftmost curve ($I \approx 0$), MI is only approximately zero because of the finite bin size, and the peak actually sharpens as the bin size decreases.
The simulation parameters are the same as in Fig.~\ref{fig:PDF}(b) and (c).
}
\label{fig:cond}
\end{figure}
Before proceeding, we mention the following points:
First, our ensemble of $\left( a_i, b_i, c_i, d_i \right)$ 
yields flat distribution of $C$
[Fig.~\ref{fig:PDF}(b)]
because the coefficients are sampled from a uniform distribution on $S^3$.
Second, $p(I)$ has a maximum at $I=0$ [Fig.~\ref{fig:PDF}(c)]
because zero MI is still highly probable even if $C \neq 0$.
For example, MI becomes zero when $|a_i||d_i |= |b_i| | c_i |$ (\ref{sec:appendix}), but this result does not always imply $a_id_i=b_i c_i$, for which $C=0$.
Figure~\ref{fig:PDF}(a) indicates that the probability of having $I=0$ is actually high regardless of $C$, but this is of little practical importance here, because it does not provide any useful insight to relate $C$ and $I$.

\begin{table}[h]
\caption{Statistical quantities
obtained from $p(C|I)$ at each given $I$.
We have calculated the peak position $C^*(I)$, the mean value $\langle C \rangle_I \equiv \int dC~C~p(C|I)$, 
and the standard deviation $\sigma_{C|I} \equiv \left[ \langle C^2 
\rangle_I - \langle C \rangle_I^2 \right]^{1/2}$.
Numbers in parentheses: Inverse of $I = \mathcal{I}(C)$ [Eq.~\eref{eq:bound}].
Simulation parameters are the same as in Fig.~\ref{fig:cond};
all entries in this table have been rounded to two decimal places.
}
\centering
\resizebox{0.4\textwidth}{!}{%
\begin{tabular}{|l|l|l|l|}
\hline
\,\,$I$              & \,$C^*$\, ($\mathcal{I}^{-1}$) & $\langle C \rangle_I$ & $\sigma_{C|I}$  \\ \hline
0.00\,           & 0.02\, (0.04)\,    & 0.23\,   & 0.28\, \\ \hline
0.10\,           & 0.37\, (0.37)\,    & 0.51\,   & 0.19\, \\ \hline
0.20\,           & 0.52\, (0.52)\,    & 0.62\,   & 0.16\, \\ \hline
0.30\,           & 0.62\, (0.62)\,    & 0.69\,   & 0.13\, \\ \hline
0.40\,           & 0.71\, (0.71)\,    & 0.76\,   & 0.11\, \\ \hline
0.50\,           & 0.78\, (0.78)\,    & 0.81\,   & 0.09\, \\ \hline 
0.60\,           & 0.84\, (0.84)\,    & 0.86\,   & 0.07\, \\ \hline
0.70\,           & 0.89\, (0.89)\,    & 0.90\,   & 0.05\, \\ \hline
0.80\,           & 0.94\, (0.94)\,    & 0.94\,   & 0.03\, \\ \hline
0.90\,           & 0.97\, (0.97)\,    & 0.97\,   & 0.02\, \\ \hline
\end{tabular}%
}
\label{tab:statistics}
\end{table}
To clarify the meaning of the nontrivial correlation,
we can check two conditional PDFs, defined as
$p(I|C) \equiv p(C,I)/p(C) $ and $p(C|I) \equiv p(C,I)/p(I)$.
$p(I|C)$ generally has two peaks,
one at $I=0$ as mentioned above, and the other at $I>0$ [Fig.~\ref{fig:cond}(a)],
whereas $p(C|I)$ has a single peak, so one can readily
infer the amount of entanglement from the observed value of $I$ in the ensemble
(Table~\ref{tab:statistics}).

\section{Discussion}
\label{sec:partial}

In this section, we will discuss how to pinpoint the peak position of $p(C|I)$ analytically.
A point in a four-dimensional real space can be represented by a pair of
complex numbers such as
$(z_1, z_2)=(x_1 + {\rm i} y_1, x_2 + {\rm i} y_2) $, where 
$x_1, x_2, y_1$, and $y_2$
are real numbers. If the complex numbers are represented in polar form, i.e.,
$z_1 = A e^{{\rm i} \alpha}$ and  $z_2 = B e^{{\rm i} \beta}$ with
$A \equiv |z_1|$ and $B \equiv |z_2|$,
a pure state can be written as
\begin{equation}
\vert \psi \rangle =  A\cos \alpha \vert 00 \rangle 
+ A\sin \alpha \vert 01\rangle 
+ B\cos \beta \vert 10 \rangle 
+  B\sin \beta \vert 11\rangle \,,
\label{eq:sincos}
\end{equation}
where $A^2 + B^2 =1$. If the coefficients of $\vert \psi \rangle$ have a
uniform random distribution on $S^3$,
all of $\alpha$, $\beta$, and $y \equiv A^2$ are uniform random variables~\cite{marsaglia1972choosing}.
Plugging Eq.~\eref{eq:sincos} into Eq.~\eref{eq:C} yields
an expression for concurrence as follows:
\begin{equation}
C = 2AB\sqrt{ (\cos\alpha \sin \beta -\sin\alpha \cos\beta)^2 }  
= 2AB| \sin \Delta| \,,
\label{eq:C2}
\end{equation}
where $\Delta \equiv \alpha-\beta$.
When $\Delta$ is given, $C$ can be obtained as a function of $y$:
\begin{equation}
    C(y)=2\sqrt{y(1-y)} |\sin\Delta|.
\end{equation}
The distribution of $y$ is given by construction, so we can obtain the PDF of $C$ in the following way:
\begin{equation}
    p (C|\Delta) = u(y) \left| \frac{dy}{dC} \right|,
    \label{eq:density}
\end{equation}
where $p(C|\Delta)$ is the conditional PDF of $C$ for given $\Delta$, and $u(y)$ is the uniform PDF of $y$.
The phase variable $\Delta$ also follows a uniform PDF $u(\Delta)$, and
one can prove that $p(C) =  \int d\Delta p(C|\Delta) u(\Delta)=1$ [Fig.~\ref{fig:PDF}(b)].
The conditional PDF on the left-hand side of Eq.~\eref{eq:density}
has a peak at $y= 1/2$ because $dC/dy$ vanishes there. For this reason, we may focus on a subset of the Hilbert space in which concurrence is simply given as
\begin{equation}
    C = \left| \sin \Delta \right|.
\end{equation}
At the same time, by setting $A = B = 1/\sqrt{2}$, we have $H_{\mathrm{L}}=1$ because $P
_1+P_2 = P_3+P_4 = 1/2$ [see Eq.~\eref{eq:hl}]. 
The value of $H_{\mathrm{L}}$ is fixed because of our parametrization in Eq.~\eref{eq:sincos}: If we had exchanged the second and the third coefficients in Eq.~\eref{eq:sincos}, we would have found $H_{\mathrm{R}}=1$ fixed instead.
\begin{figure}[h]
\centering
\includegraphics[width=0.8\textwidth]{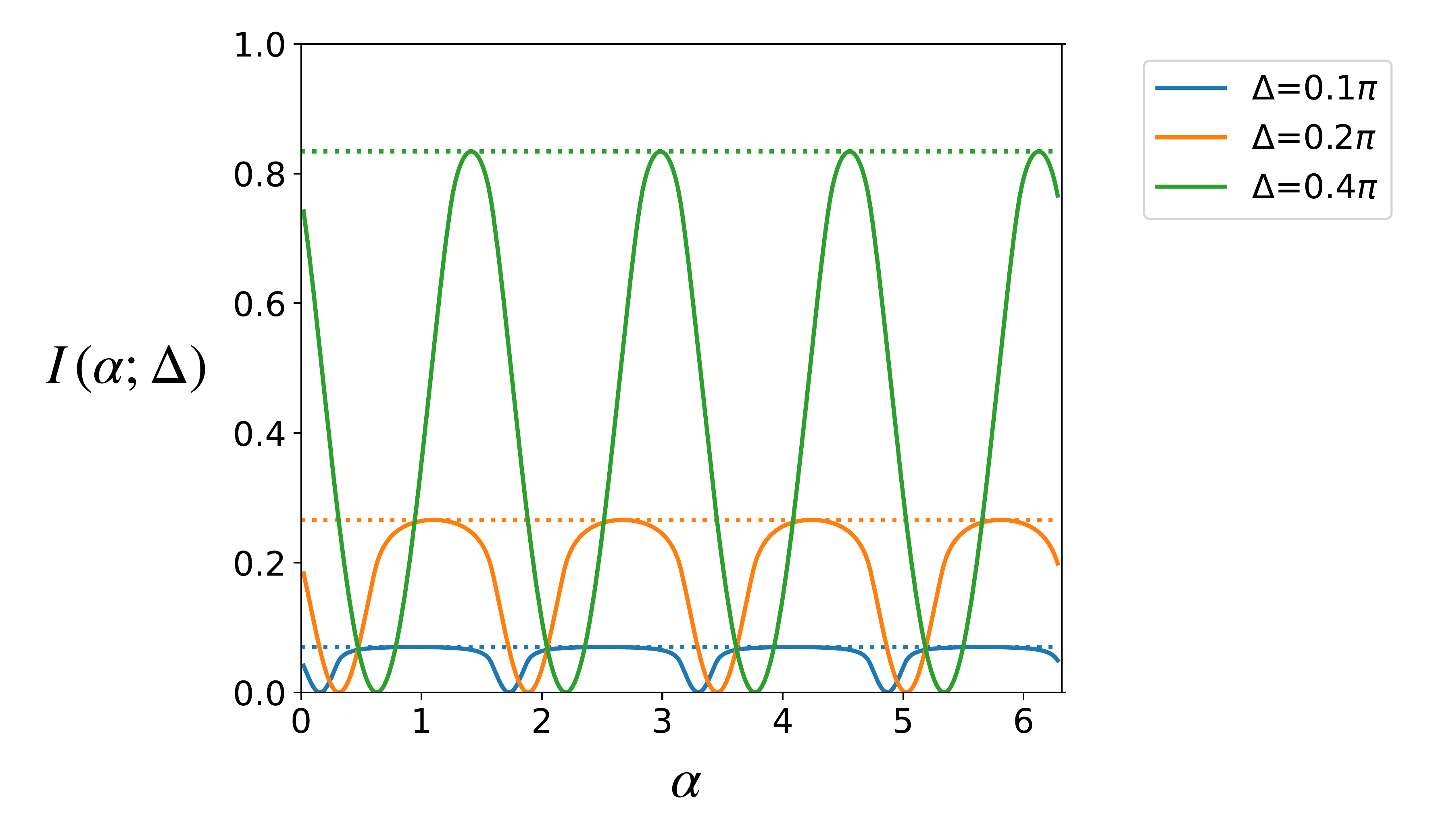}
\caption{Mutual information $I(\alpha; \Delta)$ [Eq.~\eref{eq:I2}]
as a function of $\alpha$ at different values of $\Delta \equiv \alpha-\beta$.
We have chosen $\Delta=0.1 \pi$, $0.2 \pi$, and $0.4 \pi$ to obtain the curves.
Each horizontal dotted line shows the maximum of $I$ [Eq.~\eref{eq:bound}].
}
\label{fig:periodic}
\end{figure}
Now, we can calculate MI between L and R as 
a function of $\alpha$ and $\beta$. If we eliminate $\beta$ by using $\beta = \alpha - \Delta$, MI is written as
\begin{eqnarray}
I \left( \alpha; \Delta \right)
&=& -\frac{\cos^2 \alpha  + \cos^2 \left(\alpha -\Delta \right) }{2}
\log_2 \frac{\cos^2 \alpha  + \cos^2 \left(\alpha -\Delta \right) }{2} 
\nonumber\\
&&- \frac{\sin^2 \alpha  + \sin^2 \left(\alpha -\Delta \right) }{2}
\log_2 \frac{\sin^2 \alpha + \sin^2 \left(\alpha -\Delta \right) }{2} \,\nonumber\\
&& +\frac{1}{2} \left[ \cos^2 \alpha \log_2 \cos^2 \alpha
+ \cos^2 \left(\alpha -\Delta \right) \log_2 
\cos^2 \left(\alpha -\Delta \right) \right. \nonumber\\
&&\left. +\sin^2 \alpha \log_2 \sin^2 \alpha
+ \sin^2 \left(\alpha -\Delta \right) \log_2 
\sin^2 \left(\alpha-\Delta \right) \right]\,,
\label{eq:I2}
\end{eqnarray}
which is a periodic function of $\alpha$ (Fig.~\ref{fig:periodic}).
The derivative of $I(\alpha ;\Delta)$ with respect to $\alpha$
vanishes at $\alpha^{\rm max}_n \equiv \left[\Delta + (n+1/2) \pi \right] /2$ and $\alpha^{\rm min}_n = \left( \Delta + n\pi \right)/2$, where $n$ is an integer. The vanishing derivative implies that the PDF of $I$ will peak there; this conclusion can also be argued in a similar way to Eq.~\eref{eq:density}.
At $\alpha = \alpha^{\rm min}_n$, MI $I=0$ is a minimum, and this result explains why $I=0$ is observed with high probability in Fig.~\ref{fig:PDF}(a).
Another peak position in the density of states is $\alpha = \alpha^{\rm max}_n$, at which the maximum value $I(\alpha^{\rm max}_n; \Delta)$ equals
\begin{equation}
\mathcal{I}(C) = 1 +\frac{1+C}{2} \log_2 \frac{1+C}{2}
+\frac{1-C}{2} \log_2 \frac{1-C}{2}\,.
\label{eq:bound}
\end{equation}
In this way, we predict one of the most probable values of MI.
Equation~\eref{eq:bound} indeed explains the narrow band in $p(C,I)$ [Fig.~\ref{fig:PDF}(a), open circles].
We also note that $H_{\mathrm{R}} = 1$ at $\alpha = \alpha^{\rm max}_n$,
so that the system has left-right symmetry when MI is maximized.
In contrast, $H_{\mathrm{R}}$ reaches $E(C)$ at
$\alpha = \alpha^{\rm min}_n$ where MI vanishes ($I=0$).

\section{Summary}
\label{sec:summary}
We have investigated the correlation between concurrence 
and classical MI by calculating their joint and conditional 
PDFs in an ensemble of random two-qubit pure states.
MI depends on the measurement basis, and
we have considered the product of the local basis states,
which should be most feasible experimentally.
Although MI is a poor measure of QE in general,
we have found that the PDFs have nontrivial structures:
Between the general upper bound $I=E(C)$ and the trivial lower bound $I=0$,
a nontrivial peak exists at $I = \mathcal{I}(C)$ [Eq.~\eref{eq:bound}].
By using this correlation between entanglement and post-measurement MI,
one can statistically infer the amount of entanglement through classical processes.

We stress that we have chosen uniform distribution of pure states only for the sake of analytic convenience,
and that our main argument that considers singularity in the density of states is largely insensitive to the specific distribution of the coefficients.
However, a different ensemble such as thermal equilibrium may yield a different correlation pattern, and this would be an important direction from the aspect of real applications.

\ack
Y.K. and S.K.B. were supported by Basic Science Research Program through the
National Research Foundation of Korea (NRF) funded by the Ministry of
Education (Grant No. NRF-2020R1I1A2071670).
J.U. was supported by the NRF Grant No.~2020R1I1A1A01071924.

\appendix
\section{Mutual information when $|a_i||d_i|=|b_i||c_i|$}
\label{sec:appendix}
MI in Eq.~\eref{eq:I} is written in terms of 
real numbers, $a_i$, $b_i$, $c_i$, and $d_i$ as
\begin{eqnarray}
I[\psi_i]&=&-\left( a_i^2 + b_i^2 \right) \ln a_i^2 \left( 1 + \frac{b_i^2}{a_i^2} \right)
-\left( c_i^2 + d_i^2 \right) \ln c_i^2 \left( 1 + \frac{d_i^2}{c_i^2} \right) \nonumber\\
&& -\left( a_i^2 + c_i^2 \right) \ln a_i^2\left( 1 + \frac{c_i^2}{a_i^2} \right)
-\left( b_i^2 + d_i^2 \right) \ln b_i^2 \left( 1 + \frac{d_i^2}{b_i^2} \right) \nonumber\\
&&~~~+a_i^2 \ln a_i^2 + b_i^2 \ln b_i^2 + c_i^2 \ln c_i^2 + d_i^2 \ln d_i^2 \,.
\end{eqnarray}
The condition that $|a_i||d_i|=|b_i||c_i|$,
is equivalent to $a_i^2d_i^2=b_i^2c_i^2$, which yields
$1+b_i^2/a_i^2 = 1+ d_i^2/c_i^2$ and $1+c_i^2/a_i^2 = 1+ b_i^2/d_i^2$. MI $I[\psi_i]$ then equals
\begin{equation}
I[\psi_i]= -\ln\left( 1+ \frac{b_i^2}{a_i^2} \right) \left( 1+ \frac{c_i^2}{a_i^2} \right)  
-\ln a_i^2 -d_i^2  \ln \frac{b_i^2 c_i^2}{a_i^2d_i^2}\,,
\label{eq:zeroI}
\end{equation}
where the last term vanishes. In fact, the first and second terms in Eq.~\eref{eq:zeroI} 
cancel out each other because
\begin{equation}
\left( 1+ \frac{b_i^2}{a_i^2} \right) \left( 1+ \frac{c_i^2}{a_i^2} \right)
= 1 + \frac{b_i^2}{a_i^2} +\frac{c_i^2}{a_i^2} + \frac{d_i^2}{a_i^2} = \frac{1}{a_i^2}\,, 
\end{equation}
where we have used $b_i^2c_i^2/a_i^4 = d_i^2/a_i^2$.

\section*{References}
\bibliographystyle{iopart-num}
\bibliography{cc}

\end{document}